\documentstyle[twocolumn,twoside]{article}
\newcommand{\hb}{\\ \hspace*{2ex}}

\begin{document}
\title{INTERACTION OF SUPERNOVA BLAST WAVES WITH WIND-DRIVEN
SHELLS: FORMATION OF "JETS", "BULLETS", "EARS", ETC}
\author{V.V.\,Gvaramadze$^{1,2,3}$\\[2mm]
\begin{tabular}{l}
 $^1$ Abastumani Astrophysical Observatory, Georgian Academy of Sciences,\hb
 A.Kazbegi ave. 2-a, Tbilisi, 380060, Georgia, {\em vgvaram@mx.iki.rssi.ru}\\
 $^2$ Sternberg State Astronomical Institute, Moscow, Russia\\
$^3$ Abdus Salam International Centre for Theoretical Physics,
Trieste, Italy\\[2mm]
\end{tabular}
}
\date{}
\maketitle

ABSTRACT.
Most of middle-aged supernova remnants (SNRs) have a distorted
and complicated appearance which cannot be explained in the
framework of the Sedov-Taylor model. We consider three typical
examples of such SNRs (Vela SNR, MSH\,15-52, G\,309.2-00.6) and
show that their structure could be explained as a result of
interaction of a supernova (SN) blast wave with the ambient
medium preprocessed by the action of the SN progenitor's wind and
ionized emission.\\[1mm]

{\bf 1. Introduction}\\[1mm]

Most of middle-aged SNRs have a distorted and complicated
appearance which cannot be explained in the framework of the
standard Sedov-Taylor model. Three possibilities are usually
considered to describe the general structure of such remnants:\\
-- the SN blast wave interacts with the
inhomogeneous (density stratified and/or clumpy) interstellar
medium;\\
-- the SN ejecta is anisotropic and/or clumpy;\\ 
-- the stellar remnant (e.g. a pulsar) is a source of the
relativistic wind and/or collimated outflows (jets) which
power the central synchrotron nebula (plerion) and/or interact
with the SNR's shell.\\

For example, all above possibilities were considered to explain the
structure of the Vela SNR. Namely, the general asymmetry of this
remnant (the northeast half of the Vela SNR faced towards the
Galactic plane has a nearly circular boundary, whereas the
opposite half is very distorted) as well as its patchy
appearance in soft X-rays were attributed to the expansion of
the SN blast wave in the inhomogeneous (large-scale cloud + a
multitude of cloudlets) interstellar medium (e.g. Kahn et al.
1985, Bocchino et al. 1997). One of consequences of this
suggestion is the proposal that the origin of optical filaments
constituting the shell of the remnant is due to the slowing
and cooling of parts of the SN blast wave propagating into 
dense clumps of matter (cloudlets). A number of radial
structures (most prominent in soft X-rays) protruding far
outside the main body of the remnant was interpreted as bow shocks
produced by fragments of the exploded SN star ("bullets")
supersonically moving through the interstellar medium
(Aschenbach et al. 1995). An elongated X-ray
structure stretched from the Vela pulsar position to the center
of the brightest radio component of the Vela SNR (known as
Vela\,X) was interpreted as a one-sided jet emanating from the
Vela pulsar and transferring the pulsar's slow-down energy to
the Vela\,X (e.g. Markwardt \& \"Ogelman 1995). This
interpretation supports the proposal of Weiler \& Panagia (1980)
that the Vela\,X is a plerion. A nebula of hard X-ray (2.5-10
keV) emission stretched nearly symmetrically for about
$1^{\circ}$ on either side of the pulsar in the
northeast-southwest direction was also interpreted as a plerion 
(Willmore et al. 1992). 

The first and third possibilities were considered in connection
with the SNR
MSH\,15-52 (G\,320.4-01.2). The radio map of this remnant given
by Caswell et al. (1981) shows the elongated shell consisting of
two bright components stretched parallel to the Galactic plane
and separated by a gap of weak emission. The brightest X-ray
emission of this remnant comes from two components, one of which
centres on the position of the pulsar PSR B1509-58
(located close to the geometrical center of 
the MSH\,1509-58), while the second one
coincides with the maximum of emission of the brightest (closer
to the Galactic plane) radio component and
with the bright optical nebula (known as RCW\,89). It was
suggested that the central X-ray component of the MSH\,15-52 is a
plerion (e.g. Seward et al. 1984) and that the general structure
of this remnant is affected by one (Tamura et al. 1996, Brazier
\& Becker 1997) or two (Manchester 1987, Gaensler et al. 1999) 
jets emanating from the pulsar.

And the third example is the SNR G\,309.2-00.6, which consists (at
radio wavelengths) of a nearly circular shell and two "ears" --
arclike filamentary
structures protruding from the shell in the opposite directions
(nearly parallel to the Galactic plane). It was suggested, by
analogy with the well-known system SS433/W50, that the distorted
appearance of the G\,309.2-00.6 is due to the interaction between a
pair of jets produced by the central (unvisible) stellar remnant
and the originally spherical shell of the SNR (Gaensler et al.
1998).  It was also suggested that one of linear filaments in
the northeast "ear" represents one of the proposed jets.

The goal of this paper is to show that the structure of at least
three above-mentioned SNRs could be explained as a result of
interaction of a SN blast wave with the ambient medium
preprocessed by the action of the SN progenitor's wind and
ionized emission.
\\[2mm]

{\bf 2. Interaction of SN blast waves with wind-driven shells}\\[1mm]

It is known that progenitors of most of SN stars are massive
ones (e.g. van den Bergh \& Tammann 1991). Such stars are
sources of intense stellar winds and ionizing emission which
strongly modify the ambient interstellar medium.
The ionizing radiation of the progenitor star creates an H\,II
region, the inner, homogenized part of which gradually expands
due to the continuous photoevaporation of density
inhomogeneities in stellar environs (McKee et al. 1984). If the
mechanical luminosity of the stellar wind $L$ is much smaller
than some characteristic wind luminosity, $L^{\ast} \simeq
10^{34} (S_{46} ^2 /n)^{1/3} \,\, {\rm ergs}\,{\rm s}^{-1}$,
where $S_{46}$ is the stellar ionizing flux in units of $10^{46}
\,\, {\rm photons}\,{\rm s}^{-1}$ and $n$ is the mean density
the ambient medium would have if were homogenized, the stellar
wind flows through a homogeneous medium and creates a bubble of
radius (e.g. Weaver et al. 1977) $R(t)=11L_{34} ^{1/5} n^{-1/5}
t_{6} ^{3/5}$ pc, where $L_{34} =L/(10^{34} \,{\rm
ergs}\,{\rm s}^{-1}), t_6 =t/(10^6 {\rm years})$. Initially the
expanding bubble
is surrounded by a thin, dense shell of swept-up interstellar
gas, but eventually the gas pressure in the bubble
becomes comparable to that of the ambient medium, and the bubble
stalls, while the shell disappears. The radius of the stalled
bubble is $R_{\rm s} =5.5L_{34} ^{1/2} n^{-1/2}$ pc. Since the
star continues to supply the energy in the bubble, the radius of
the bubble continues to grow, $\propto t^{1/3}$, until the
radiative losses in the bubble interior becomes comparable to
$L$. Then the bubble recedes to some stable radius $R_{\rm r}$,
at which radiative losses exactly balance L (D'Ercole 1992):
$R_{\rm r} = 2.2\,L_{34} ^{6/13} \, n^{-7/13}$ pc.
Before a massive star
exploded as a supernova it becomes for a relatively short time,
$t_{\rm RSG} \simeq 10^6$ years, a
red supergiant (RSG). The ionized gas outside the bubble
rapidly cools off because the central star cannot keep it hot. At
the same time the rarefied interior of the bubble remains hot as
the radiative losses there are negligible on time-scales of 
$t_{\rm RSG}$. As a result, the bubble
supersonically reexpands in the external cold medium and creates
a new dense shell (D'Ercole 1992; cf. Shull et al. 1985). 
Two main factors could significantly
affect the structure of the shell. The first one is the regular 
interstellar magnetic field (generally it is parallel to the Galactic
plane). This factor leads to the matter redistribution over the shell and to
its concentration near the magnetic equator: the column density
at the equator is increased about ten times (Ferri\`ere et al.
1991). The second factor is the large-scale
density gradient. It is known (Landecker et al. 1989, Gosachinskij \&
Morozova 1999) that molecular clouds tend to be stretched
along the Galactic plane, therefore one might expect that due 
to the interaction with a nearby cloud
one of two sides of the shell (not necessary the nearest to the
Galactic plane) could be more massive than the opposite one. 
These two factors naturally define two symmetry axes (parallel
and perpendicular to the Galactic plane) of the future SNR.

During the RSG stage a massive star lost most of its
mass (e.g. a $20 M_{\odot}$ star loses about two thirds of its
mass) in the form of slow, dense wind. This material
expands in the interior of the reexpanded main-sequence (MS) bubble
and occupies a compact region surrounded by a dense shell. 
The size of this region is determined by
the counter-pressure of the external hot gas and is equal to about
few parsecs (e.g. Chevalier \& Emmering 1989, D'Ercole 1992).
Most probably that this region is far from the spherical
symmetry (it is believed that the wind of a RSG is
concentrated close to the stellar equatorial plane). 

After the SN exploded, the blast wave interacts with the dense
RSG wind. This interaction continues few hundreds years and
determines the appearance of young SNRs (e.g. Cas\,A, see
Borkowski et al. 1996). Then the blast wave
propagates through the low-density interior of the MS 
bubble until it catches up the dense shell. During this
period (lasting about one thousand years) the
blast wave is unobservable. The subsequent evolution of the
blast wave (i.e. the SNR) depends on the mass of the shell. 
If the mass of the shell is smaller than about 50
times the mass of the SN ejecta the blast wave overruns the
shell and continues to expand adiabatically as a Sedov-Taylor
shock wave. For more massive ones, the blast wave merges with
the shell, and the reaccelerated shell evolves into a
momentum-conserving stage (e.g. Franco et al. 1991).
The impact of the blast wave with the shell causes the  
Rayleigh-Taylor and other dynamical instabilities. The
inhomogeneous mass distribution over the shell affects the
development of instabilities and results in the 
asymmetry of the resulting SNR. 
The more massive half of a shell created in the density-stratified
medium is less sensitive to the impact of the SN blast wave, while
the opposite (less massive) one becomes strongly deformed and
sometimes even disrupted. The effect of the regular
magnetic field is twofold: first, it leads to the bilateral 
appearance of SNRs (cf. Ferri\`ere et al. 1991, Gaensler
1998), second, it results in the elongated form of remnants
(because of reduced inertia of shells at the magnetic poles).
\\[2mm]

{\bf 3. Three examples}\\[1mm]

Let us consider the SNRs mentioned in Sect. 1.\\[2mm]

{\it 3.1. Vela SNR}\\[1mm]

We suggest that the Vela SNR is a result of type II SN explosion
in a cavity created by the wind of a 15-20 $M_{\odot}$ star and
propose that the general structure of the remnant is determined
by the interaction of the SN blast wave with the massive shell
created around the reexpanded MS bubble 
(see Sect. 2; for details see Gvaramadze (1999a)). 
The impact of the blast wave with the shell
causes the development of Rayleigh-Taylor deformations of the
shell ("blisters"), which appear as arclike and looplike
filaments when our line of sight is tangential to their
surfaces. The optical emission is expected to come from the
outer layers of the shell, where the transmitted SN blast wave
slows to become radiative, while the soft X-ray emission
represents the inner layers of the shell heated by the blast
wave up to X-ray temperatures. 
The origin of some radial protrusions (labelled by
Aschenbach et al. (1995) as "bullets" A,B,C, and D/D') could be
connected with the shell deformations, while the "bullets" E and
F could be interpreted as outflows of a hot gas escaping through
the breaks in the SNR's shell (Gvaramadze 1998a, Bock \&
Gvaramadze 1999). As to the X-ray "jet" discovered by Markwardt
\& \"Ogelman (1995), an analysis of the radio, optical, and
X-ray data suggested that it is a dense filament in the Vela
SNR's shell (projected by chance near the line of sight to the
Vela pulsar),
and that its origin is connected with the nonlinear interaction
of the shell deformations (see Gvaramadze 1999a). The nature of
the radio source Vela\,X is considered in the paper by
Gvaramadze (1998b), where it is shown that the Vela\,X is also a
part of the shell of the Vela SNR, but not a plerion. In
conclusion one should be noted that the slow, dense RSG wind
lost by the progenitor star and subsequently reheated and
reaccelerated by the passage of the SN blast wave
could be responsible for the origin of a hard X-ray nebula
discovered by Willmore et al. (1992) (Willmore et al. mentioned
that their data do not allow to discern the thermal and
nonthermal forms of the spectrum of this nebula).
\\[2mm]

{\it 3.2. SNR MSH\,15-52}\\[1mm]

The SNR MSH\,15-52, associated with the pulsar PSR B1509-58, is
usually classified as a composite SNR. This is because of it
consists of an extended nonthermal radio shell (at the distance
of $\simeq 5$ pc (e.g. Gaensler et al. 1999) the diameter of the
remnant $\geq 40$ pc) and a central elongated X-ray nebula
($\simeq 7$ pc $\times 12$ pc) which is thought to be a
synchrotron pulsar-powered nebula (a plerion). The spin-down age
of the pulsar is $\simeq 1700$ years (i.e. nearly the same as
that of the Crab pulsar), 
while the size and general appearance of the
MSH\,15-52 suggest that this system should be much older (few
times $10^4$ years). To reconcile the ages of the pulsar and
remnant, Seward et al. (1983) considered two possibilities: 1)
MSH\,15-52 is a young SNR, and 2) PSR B1509-58 is an old pulsar.
The first one implies (in the framework of the Sedov-Taylor
model) that the SN explosion was very energetic and occured in a
tenuous medium (see also Bhattacharya 1990). This point of view
is generally accepted (e.g. Gaensler et al. 1999). The second
possibility was reexaminated by Blandford \& Romani (1988), who
suggested that the pulsar spin-down torque grew within the last
$\simeq 10^3$ years (due to the growth of the pulsar's magnetic
field) and therefore the true age of the pulsar
could be as large as it follows from the age estimates for the
SNR. We propose an alternative explanation (Gvaramadze 1999b,c)
and suggest that the high spin-down rate of the pulsar is
inherent only for a relatively short period of the present spin
history and that the enhanced braking torque is connected with
the interaction of the pulsar's
magnetosphere with a dense clump of circumstellar matter (whose
origin is connected with the late evolutionary stages of the
progenitor star). This suggestion implies that the central X-ray
nebula could be interpreted as a dense material lost by the
progenitor star during the RSG stage and reheated to high
temperatures by the SN blast wave. The existance of a hot
plasma (of mass of about few $M_{\odot}$) around the pulsar
follows from the IR observations of the MSH\,15-52 by Arendt (1991),
who discovered an IR source near the position of the pulsar.
We believe that the thermal
emission of this plasma is contaminated by the hard nonthermal
emission from a (much smaller) compact nebula powered by the
pulsar (similar to the $1^{'} (\simeq 4\times 10^{17} \, 
{\rm cm})$ nebula discovered by Harnden
et al. (1985; see also de Jager et al. 1996) around the Vela
pulsar), and that this is the reason why the spectrum of the
whole central nebula is usually described by a nonthermal model
(e.g. Greiveldinger et al. 1995, Tamura et al. 1996). 

The shell of the MSH\,15-52 remainds that
of the Vela SNR (cf. Fig.8 of Gaensler 1998 and
Fig.1 of Gvaramadze 1999a). In both remnants the halves faced
towards the Galactic plane are brighter and more regular than
the opposite ones. We suggest that the MSH\,15-52 is a
result of interaction of the SN blast wave with the wind-driven
shell created in the inhomogeneous interstellar medium:
the northwest half of the shell interacts with the
region of enhanced density (that results in the origin of
bright radio, optical and X-ray emission), and therefore is less
affected (distorted) by the impact of the SN blast wave than the
southeast half. The bilateral and elongated appearance of the
shell could be connected with the effect of the large-scale 
interstellar magnetic field (cf. Gaensler 1998, Gaensler et al.
1999).  
\\[2mm]

{\it 3.3. SNR G\,309.2-00.6}\\[1mm]

We suggest that the "ears" of this SNR were blown up in the polar
regions of the (former) wind-driven shell created in the
interstellar medium with regular magnetic field (oriented nearly
parallel to the Galactic plane). The origin of the "jet" and
other filamentary structures visible in the remnant (see Fig.2
of Gaensler et al. 1998) we connect with projection effects in
the Rayleigh-Taylor unstable shell. We suggest also that the
SN explosion site\footnote{Note that it could be shifted from
the geometrical centre of the SNR due to the proper motion of
the SN progenitor star.} should be marked by a hard X-ray nebula and
predict that the angular size of the nebula (for the distance to
the remnant of 5-14 kpc (Gaensler et al. 1998)) is about
$1.5^{'} - 2^{'}$.
\\[2mm]

\indent
{\bf References\\[2mm]}
Arendt R.G.: 1991, {\it AJ}, {\bf 101}, 2160.\\ 
Aschenbach B., Egger R., Tr\"umper J.: 1995, {\it Nat}, {\bf
373}, 587.\\
Bhattacharya D.: 1990, {\it JA\&A}, {\bf 11}, 125.\\ 
Blandford R.D., Romani R.W.: 1988, {\it MNRAS}, {\bf 234}, 57p.\\ 
Bocchino F., Maggio A., Sciortino S.: 1997, {\it ApJ}, {\bf
481}, 872.\\ 
Bock D.C.-J., Gvaramadze V.V.: 1999, in preparation.\\
Borkowski K.J., Szymkowiak A.E., Blondin J.M., Sarazin C.L.:
1996, {\it ApJ}, {\bf 466}, 866.\\  
Brazier K.T.S., Becker W.: 1997, {\it MNRAS}, {\bf 284}, 335.\\ 
Caswell J.L., Milne D.K., Wellington K.J.: 1981, {\it MNRAS},
{\bf 195}, 89.\\ 
Chevalier R.A., Emmering R.T.: 1989, {\it ApJ}, {\bf 342}, L75.\\
D'Ercole A.: 1992, {\it MNRAS}, {\bf 255}, 572.\\
de Jager O.C., Harding A.K., Strickman M.S.: 1996, {\it ApJ},
{\bf 460}, 729.\\  
Ferri\`ere K.M., Mac Low M.-M., Zweibel E.G.: 1991, {\it ApJ},
{\bf 375}, 239.\\ 
Franco J., Tenorio-Tagle G., Bodenheimer P., R\'{o}\.{z}yczka
M.: 1991, {\it PASP}, {\bf 103}, 803.\\
Gaensler B.M.: 1998, {\it ApJ}, {\bf 493}, 781.\\
Gaensler B.M., Green A.J., Manchester R.N.: 1998, {\it MNRAS},
{\bf 299}, 812.\\ 
Gaensler B.M., Brazier K.T.S., Manchester R.N., Johnston S.,
Green A.J.: 1999, {\it MNRAS}, {\bf 305}, 724.\\
Greiveldinger C., Caucino S., Massaglia S., \"Ogelman H.,
Trussoni E.: 1995, {\it ApJ}, {\bf 454}, 855.\\ 
Gvaramadze V.V.: 1998a, in: {\it The Local Bubble and Beyond},
eds. D.Breitschwerdt, M.Freyberg, J.Tr\"umper, Springer-Verlag,
Heidelberg, p. 141.\\
Gvaramadze V.V.: 1998b, {\it Astronomy Letters}, {\bf 24}, 178.\\
Gvaramadze V.V.: 1999a, {\it A\&A}, {\bf 352}, 712.\\
Gvaramadze V.V.: 1999b, in: {\it Proceedings of the All-Russian
Conference "Astrophysics on the Boundary of Centuries" (17-22
May 1999, Pushchino, Russia)}, in press.\\
Gvaramadze V.V.: 1999c, submitted to {\it A\&A}.\\
Gosachinskij I.V., Morozova V.V.: 1999, {\it Astronomy Reports},
in press.\\
Harnden F.R., Grant P.D., Seward F.D., Kahn S.M.: 1985, {\it
ApJ}, {\bf 299}, 828.\\  
Kahn S.M., Gorenstein P., Harnden F.R., Seward F.D.: 1985, {\it
ApJ}, {\bf 299}, 821.\\ 
Landecker T.L., Pineault S., Routledge D., Vaneldik J.F.: 1989,
{\it MNRAS}, {\bf 237}, 277.\\
McKee C.F., Van Buren D., Lazareff R.: 1984, {\it ApJ}, {\bf 278}, L115.\\ 
Manchester R.N.: 1987, {\it A\&A}, {\bf 171}, 205.\\ 
Markwardt C.B., \"Ogelman H.: 1995, {\it Nat}, {\bf 375}, 40.\\
Seward F.D., Harnden Jr., F.R., Murdin P., Clark D.H.: 1983,
{\it ApJ}, {\bf 267}, 698.\\ 
Seward F.D., Harnden Jr., F.R., Szymkowiak A., Swank J.: 1984,
{\it ApJ}, {\bf 281}, 650.\\ 
Shull P., Dyson J.E., Kahn F.D., West K.A.: 1985, {\it MNRAS},
{\bf 212}, 799.\\  
Tamura K., Kawai N., Yoshida A., Brinkmann W.: 1996, {\it PASP},
{\bf 48}, L33.\\  
van den Bergh S., Tammann G.A.: 1991, {\it ARA\&A}, {\bf 29},
363.\\ 
Weaver R., McCray R., Castor J., Shapiro P., Moore R.: 1977,
{\it ApJ}, {\bf 218}, 377.\\  
Weiler K.W., Panagia N.: 1980, {\it A\&A}, {\bf 90}, 269.\\
Willmore A.P., Eyles C.J., Skinner G.K., Watt M.P.: 1992, {\it
MNRAS}, {\bf 254}, 139.\\  
\vfill
\end{document}